\documentclass[12pt]{article}
\usepackage{graphicx}
\usepackage[cp1251]{inputenc}
\usepackage{epsfig}
\usepackage[english]{babel}

\date{}
\def\b{\begin{equation}}
\def\e{\end{equation}}
\def\bee{\begin{enumerate}}
\def\eee{\end{enumerate}}
\def\be{\begin{vmatrix}}
\def\ee{\end{vmatrix}}

\begin{document}
\setcounter{page}{1}
\bibliographystyle{unsrt2}
\pagestyle{plain}

\title{\bf{Optical properties of biased bilayer graphene due to gap parameter effects}}
\author{Bahram Maleki and Hamed Rezania\thanks{Corresponding author. Tel./fax: +98 831 427 4569., Tel: +98 831 427 4569.
E-mail: rezania.hamed@gmail.com}}
\maketitle{\centerline{Department of Physics, Razi University,
Kermanshah, Iran}
\begin{abstract}
We address the optical conductivity of undoped bilayer graphene
in the presence of a finite bias voltage at finite temperature.
The effects of gap parameter and stacking type on optical conductivity are discussed
in the context of tight binding model Hamiltonian.
  Green's function approach has been implemented
to find the behavior of optical conductivity of bilayer graphene within linear response theory.
We have found the frequency dependence of optical conductivity for
different values of gap parameter and bias voltage.
Also the dependence of optical conductivity on the temperature
 has been investigated in details.
A peak appears in the plot of optical conductivity versus frequency for different
values of temperatures and bias voltage.
 Furthermore we find the frequency position of broad peak in optical conductivity
 goes to higher values with increase of gap parameter for both bernal and simple stacked bilayer graphenes.
\end{abstract}
\vspace{0.5cm} {\it \emph{Keywords}}: bilayer Graphene; Green's function; Optical conductivity
\section{Introduction}
The isolation of monolayer graphene has generated a lot of
interesting topics for the physics community. On the one hand, the electronic
spectrum of graphene can be described by the two dimensional Dirac
equation. \cite{novoselov,novo,geim22}. However, the gapless
electron spectrum of monolayer graphene makes it difficult to turn
off the electrical current due to tunneling. Bilayer graphene , on
the other hand, can provide a finite band gap up to hundreds of meV,
when the inversion symmetry between top and bottom layers is broken
by an applied perpendicular electric field\cite{mccann1,min}. This
system is composed of two layers, each of which has carbon atoms
arranged in a honeycomb lattice with two sublattices. Each layer can
be considered by strong interlayer hopping integral between two
different sublattices sites in the two layers. One can consider a
simple model for bilayer graphene spectrum is a pair of chiral
parabolic electron and hole bands touching each other at the Dirac
point.

In contrast to the case of single-layer graphene (SLG)
low energy excitations of the bilayer graphene
have parabolic spectrum, although, the chiral form of the effective
2-band hamiltonian persists
because the sublattice pseudospin is still a relevant degree of freedom.
The pristine or undoped bilayer graphene has attracted a lot of interest since the presence
of a single Fermi point and quadratic dispersion can lead to a host of exotic phenomena
\cite{barlas,levitov}.

The low energy approximation
in bilayer graphene is valid only for small doping $n<10^{12}cm^{-2}$,
while experimentally doping
can obtain 10 times larger densities. For such a large doping, the 4-band model\cite{mccann}
should be used
instead of the low energy effective 2-band model.
Furthermore, an electronic band gap can be introduced in a dual gate bilayer graphene
\cite{mccann,lopes}, and it makes BLG very appealing from the point of view of applications.
It was shown theoretically
\cite{mccann1,mccann} and
demonstrated experimentally\cite{oostinga,lopes} that a
graphene bilayer is the only material with semiconducting properties
that can be controlled by electric field effect\cite{min}.

The optical properties of graphene are of considerable importance
for technological applications and all variants of graphene are also of
potential interest and should be examined. The dynamical
conductivity of graphene has been extensively studied
theoretically\cite{ando} and experiments have largely verified the
expected behavior\cite{nair}. The conductivity for Bernal-stacked
bilayer graphene has been studied theoretically\cite{abergal} and
observed\cite{fogler}. Some preliminary work on the absorption
coefficient of undoped AA-stacked graphene in zero magnetic field
has been reported\cite{dong} however most materials naturally occur
with charge doping where the Fermi level is away from charge
neutrality. Furthermore, the interesting feature for practical
applications is the variation of optical properties with doping,
usually achieved through a field effect transistor structure. The
optical conductivity of a few-layer epitaxial graphite\cite{sadovs}
and oriented pyrolytic graphite \cite{li1} in finite external
magnetic field has been reported recently, as well as for
graphene\cite{jiang}. There have also been theoretical studies
\cite{falko,shara} of the conductivity, including discussions of
optical sum rules\cite{carbo} which continue to provide useful
information on the electron dynamics. In the other theoretical work,
frequency dependence of
 dynamical conductivity of AA-stacked bilayer graphene in the presence of electron doping has
been calculated within Dirac approximation\cite{tarbetArxiv}. For
in-plane response, AA-stacked graphene shows a Drude weight at
charge neutrality along with Pauli blocking at low frequencies below
the onset of a flat interband absorption.

In an experimental work the effects of both bias voltage and
impurity doping on Raman spectrum and electrical resistivity have
been studied\cite{han}. In this experimental study the transport
results presents a variable range hopping conduction near the charge
neutrality point at low temperatures. Such results provide evidence
for the impurity level inside the gap. Also magnetoresistance of
biased bilayer graphene as a function of magnetic field has been
studied. The electron-impurity scattering rate as a function of
quasi particle energy for doped bilayer graphene has been
theoretically calculated\cite{patel}. The results show that
scattering rate for bilayer graphene enhance on increasing impurity
concentrations. Also electron impurity scattering rate of bilayer
graphene goes to a constant value for zero limit of quasi particle
energy. The other theoretical works study the impurity levels and
the effects of impurity doping on the local density of states in the
gapped doped bilayer graphene\cite{nilsson1,koshino1,ferreira,mikhi}

In this paper, we study the effects of bias voltage and energy gap
 on the optical conductivity of both simple and bernal stacked undoped graphene bilayer within full band tight binding
approximation.
Full band calculation beyond Dirac approximation has been implemented to derive in-plane
  optical conductivity
spectra. We have exploited Green's function approach to calculate the
optical conductivity, i.e. the time ordered electrical current correlation.
 The effects of energy gap, bias voltage and stacking types on the frequency dependence of optical conductivity of bilayer graphene
 have been studied. Also we discuss and analyze to
show how bias voltage value affects the frequency dependence of the optical conductivity. We study
 the optical spectra of bilayer graphene along zigzag direction
for both types of bilayer graphene versus photon frequency. Finally we have compared the behavior of
 optical conductivity of bilayer graphene for two different stacking types in details.
 \section{Theoretical Model and Method}
To calculate the optical conductivity of bilayer graphene
we consider bilayer graphene composed of
two graphene single layers arranged in both cases of the simple (AA) and Bernal (AB) stacking
\cite{nilsson}.
This lattice structure of each graphene layer has been shown in Fig.(1).
The unit cell vectors of bilayer graphene is similar to those of single
layer one.
The
primitive unit cell vectors are given by
\begin{eqnarray}
 {\bf a}_{1}=\frac{\sqrt{3}}{2}{\bf i}+\frac{{\bf j}}{2}\;\;,\;\;{\bf a}_{2}=\frac{\sqrt{3}}{2}{\bf i}-\frac{{\bf j}}{2},
\end{eqnarray}
where ${\bf i}$ and ${\bf j}$ are unit cell vectors along zigzag and armchair directions, respectively.
Also the length of unit cell vectors is considered to be one.
In order to obtain optical properties of bilayer graphene
 we must first examine the band structure and provide and expression for the electronic
Green's function. We start from a tight
binding model incorporating nearest neighboring intralayer and interlayer
hopping terms. An on-site potential energy difference between two layers is included to model
the effect of an external voltage. Also a finite difference between on-site energies of two different sublattice atoms of honeycomb structure
has been applied in the model Hamiltonian.
For the case of AA-stacking, an A (B) atom in the upper layer is
stacked directly above A(B) atom in the lower layer.
 The spin independent tight binding model hamiltonians in the nearest neighbor approximation for
AA-stacked bilayer grapehene ($H_{AA}$) and AB one ($H_{AB}$) are given by
\begin{eqnarray}
 H_{AA}&=&-t\sum_{l=1,2}\sum_{i,\delta}(b^{\dag}_{l,i+\delta}a_{l,i}+h.c.)-
\gamma\sum_{i}a^{\dag}_{1,i}a_{2,i}+h.c.-\gamma\sum_{i}b^{\dag}_{1,i}b_{2,i}+h.c.
\nonumber\\&+&\frac{V}{2}\sum_{i}(a^{\dag}_{1,i}a_{1,i}+b^{\dag}_{1,i}b_{1,i})-
\frac{V}{2}\sum_{i}(a^{\dag}_{2,i}a_{2,i}+b^{\dag}_{2,i}b_{2,i})+
\Delta\sum_{i}(a^{\dag}_{1,i}a_{1,i}+a^{\dag}_{2,i}a_{2,i})\nonumber\\
&-&
\Delta\sum_{i}(b^{\dag}_{1,i}b_{1,i}+b^{\dag}_{2,i}b_{2,i})
,\nonumber\\
H_{AB}&=&-t\sum_{l=1,2}\sum_{i,\delta}(b^{\dag}_{l,i+\delta}a_{l,i}+h.c.)-
\gamma\sum_{i}a^{\dag}_{1,i}b_{2,i}+h.c.
\nonumber\\&+&\frac{V}{2}\sum_{i}(a^{\dag}_{1,i}a_{1,i}+b^{\dag}_{1,i}b_{1,i})-
\frac{V}{2}\sum_{i}(a^{\dag}_{2,i}a_{2,i}+b^{\dag}_{2,i}b_{2,i})+
\Delta\sum_{i}(a^{\dag}_{1,i}a_{1,i}+a^{\dag}_{2,i}a_{2,i})\nonumber\\
&-&
\Delta\sum_{i}(b^{\dag}_{1,i}b_{1,i}+b^{\dag}_{2,i}b_{2,i}).
\label{e1}
\end{eqnarray}
In Eq.(\ref{e1}), $a_{l,i}(b_{l,i})$ denotes the annihilation operator for an electron
which is on an A(B)-atom site with unit cell label $i$ in the graphene layer indexed by $l$.
 $\delta$ is one of three vectors that connects the unit cells of nearest-neighbor lattice sites and given by
$\delta=0,{\bf a}_{1},{\bf a}_{2}$ according to Fig.(1).
$t\approx 2.6 eV$ is the nearest neighbour intralayer hopping terms for electrons to
move within a given plane.
The third term in Eq.(\ref{e1}) corresponds to the interlayer hopping between graphene sheets with
amplitude $\gamma$. In bilayer graphene,
 we have
$\gamma\approx 0.2 eV$ \cite{dong,lobato}.
 $V$ refers to the potential energy difference between the first and second layers
induced by a bias voltage.\\
The following Fourier transformations for
fermionic operators $a^{\dag}_{l,i}$,$b^{\dag}_{l,i}$ have been given by
\begin{eqnarray}
 a^{\dag}_{l,{\bf k}}&=&\frac{1}{\sqrt{N}}\sum_{i}e^{-i{\bf k}.{\bf R}_{i}}a^{\dag}_{l,i},\nonumber\\
b^{\dag}_{l,{\bf k}}&=&\frac{1}{\sqrt{N}}\sum_{i}e^{-i{\bf k}.{\bf R}_{i}}b^{\dag}_{l,i},
\end{eqnarray}
where $N$ is the number of unit cells and ${\bf k}$
 is wave vector belonging to the first Brillouin zone of honeycomb structure.
${\bf R}_{i}$ introduces the position vector of $i$ th unit cell in graphene layer.
In terms of Fourier transformation of
operators, one can rewrite the clean tight binding part of
the Hamiltonians in Eq.(\ref{e1}) as
\begin{eqnarray}
 H_{AA}&=&\sum_{{\bf k}}\phi^{\dag}_{{\bf k}}H_{AA}({\bf k})\phi_{{\bf k}},\nonumber\\
 H_{AB}&=&\sum_{{\bf k}}\phi^{\dag}_{{\bf k}}H_{AB}({\bf k})\phi_{{\bf k}},
\label{e10.5.2.1}
\end{eqnarray}
in which the vector of fermion creation operators is defined as
$\phi^{\dag}_{{\bf k}}=(a^{\dag}_{1,{\bf k}},b^{\dag}_{2,{\bf k}},
a^{\dag}_{2,{\bf k}},b^{\dag}_{1,{\bf k}})$.
 The nearest neighbor
approximation gives us the following matrix forms for $H_{AA}({\bf k})$ and
$H_{AB}({\bf k})$ as
\begin{eqnarray}
H_{AA}({\bf k})
&=&\left(
                              \begin{array}{cccc}
                                \Delta+V/2 & 0 & \gamma & f({\bf k}) \\
                                0 & -\Delta-V/2& f^{\ast}({\bf k}) & \gamma \\
                                \gamma & f({\bf k}) & \Delta-V/2 & 0 \\
                                f^{\ast}({\bf k}) & \gamma & 0 & -\Delta+V/2\\
  \end{array}
\right),\nonumber\\
H_{AB}({\bf k})
&=&\left(
                              \begin{array}{cccc}
                                \Delta+V/2+g({\bf k}) & 0 & 0 & f({\bf k}) \\
                                0 & -\Delta-V/2& f^{\ast}({\bf k}) & 0 \\
                                0 & f({\bf k}) & \Delta-V/2 & \gamma \\
                                f^{\ast}({\bf k}) & 0 & \gamma & -\Delta+V/2\\
  \end{array}
\right).
 \label{a10.5.1}
\end{eqnarray}
Using the definition for $\delta$,
the function $f({\bf k})$ are expressed based on the following relations
\begin{eqnarray}
f({\bf k})&=&-t \Big(1+2cos(\frac{k_{y}}{2})exp(-i\frac{\sqrt{3}}{2}k_{x})\Big).
\label{a10.5.1.1}
\end{eqnarray}
After diagonalizing of the
Hamiltonians in Eq.(\ref{a10.5.1}), the band structures for simple stacked bilayer graphene
 are given by following eigenvalues
\begin{eqnarray}
E_{1}({\bf k})&=&\sqrt{\Delta^{2}+\frac{V^{2}}{4}+\gamma^{2}
+|f({\bf k})|^{2}+
2\sqrt{\Big(\frac{V^{2}}{4}+\gamma^{2}\Big)
\Big(|f({\bf k})|^{2}+\Delta^{2}}\Big)},\nonumber\\
E_{2}({\bf k})&=&\sqrt{\Delta^{2}+\frac{V^{2}}{4}+\gamma^{2}
+|f({\bf k})|^{2}-
2\sqrt{\Big(\frac{V^{2}}{4}+\gamma^{2}\Big)
\Big(|f({\bf k})|^{2}+\Delta^{2}}\Big)},\nonumber\\
E_{3}({\bf k})&=&-\sqrt{\Delta^{2}+\frac{V^{2}}{4}+\gamma^{2}
+|f({\bf k})|^{2}+
2\sqrt{\Big(\frac{V^{2}}{4}+\gamma^{2}\Big)
\Big(|f({\bf k})|^{2}+\Delta^{2}}\Big)},\nonumber\\
E_{4}({\bf k})&=&-\sqrt{\Delta^{2}+\frac{V^{2}}{4}+\gamma^{2}
+|f({\bf k})|^{2}-
2\sqrt{\Big(\frac{V^{2}}{4}+\gamma^{2}\Big)
\Big(|f({\bf k})|^{2}+\Delta^{2}}\Big)}
\label{e0.6}
\end{eqnarray}
Also the band structure for AB stacked bilayer graphene in the presence of
bias voltage, next
nearest neighbor hopping and gap parameter is readily obtained as following expression
\begin{eqnarray}
E_{1}({\bf k})&=&\sqrt{\Delta^{2}+\frac{V^{2}}{4}+\frac{\gamma^{2}}{2}
+|f({\bf k})|^{2}+
\sqrt{\frac{\gamma^{4}}{4}+\gamma^{2}\Big(
|f({\bf k})|^{2}-V\Delta\Big)+V^{2}(|f({\bf k})|^{2}+\Delta^{2})}},\nonumber\\
E_{2}({\bf k})&=&\sqrt{\Delta^{2}+\frac{V^{2}}{4}+\frac{\gamma^{2}}{2}
+|f({\bf k})|^{2}-
\sqrt{\frac{\gamma^{4}}{4}+\gamma^{2}\Big(
|f({\bf k})|^{2}-V\Delta\Big)+V^{2}(|f({\bf k})|^{2}+\Delta^{2})}},\nonumber\\
E_{3}({\bf k})&=&-\sqrt{\Delta^{2}+\frac{V^{2}}{4}+\frac{\gamma^{2}}{2}
+|f({\bf k})|^{2}+
\sqrt{\frac{\gamma^{4}}{4}+\gamma^{2}\Big(
|f({\bf k})|^{2}-V\Delta\Big)+V^{2}(|f({\bf k})|^{2}+\Delta^{2})}},\nonumber\\
E_{4}({\bf k})&=&-\sqrt{\Delta^{2}+\frac{V^{2}}{4}+\frac{\gamma^{2}}{2}
+|f({\bf k})|^{2}-
\sqrt{\frac{\gamma^{4}}{4}+\gamma^{2}\Big(
|f({\bf k})|^{2}-V\Delta\Big)+V^{2}(|f({\bf k})|^{2}+\Delta^{2})}}\label{e0.6.1.1}
\end{eqnarray}
Using band energy spectrum in Eq.(\ref{e0.6}),
the Hamiltonian in Eq.(\ref{e10.5.2.1}) can be rewritten by
\begin{eqnarray}
 H=\sum_{k_{x},p,\eta=1,2,3,4}
E_{\eta}({\bf k})c^{\dag}_{\eta,{\bf k}}c_{\eta,{\bf k}}.
\label{e0.57}
\end{eqnarray}
The
electronic Green's function can be defined using the Hamiltonian in Eq.(\ref{e0.57}) as following expression
\begin{eqnarray}
G_{\eta}({\bf k},\tau)=-\langle T_{\tau}
c_{\eta,{\bf k}}(\tau)c^{\dag}_{\eta,{\bf k}}(0)\rangle,
 \label{e0.601}
\end{eqnarray}
where $\tau$ is imaginary time.
Using the model Hamiltonian in Eq.(\ref{e0.57}),
the Fourier transformations of Green's function is given by
\begin{eqnarray}
 G_{\eta}({\bf k},i\omega_{n})=\int^{1/k_{B}T}_{0}d\tau e^{i\omega_{n}\tau}
G_{\eta}({\bf k},\tau)=\frac{1}{i\omega_{n}-E_{\eta}({\bf k})}.
\label{e0.61}
\end{eqnarray}
Here $\omega_{n}=(2n+1)\pi k_{B}T$ denotes the fermionic Matsubara frequency
in which $T$
is equilibrium temperature.
In the following, the relation of the real part of in-plane optical conductivity of AA stacked bilayer graphene
is presented using Green's function method\cite{mahan}. When bilayer graphene
 is excited by the electromagnetic field with polarization
of electric field along $x$ direction presented in Fig.(1),
an Hamiltonian term as $\frac{e}{mc}{\bf A}.{\bf p}$ is
added to original Hamiltonian $H=H_{0}+H_{imp}$.
The optical conductivity, as dynamical correlation function between electrical current operators, for single band model Hamiltonian
\cite{kotliar} have been obtained in terms of
Green's function using Wick's theorem.
One can generalize this result for multiband model Hamiltonian and
 the optical conductivity of bilayer graphene
along $x$ zigzag direction are related Green's function as
\begin{eqnarray}
\sigma(\omega)=\frac{e^{2}k_{B}T}{4N}\sum_{{\bf k},\eta}(\frac
{\partial E^{\sigma}_{\eta}({\bf k})}{\partial k_{x}})^{2}\int_{-\infty}^{\infty}
\frac{d\epsilon}{2\pi}\Big
(\frac{ n_{F}(\epsilon+\omega)-n_{F}(\epsilon)}{\omega}\Big)
\Big(2ImG_{\eta}({\bf k},i\omega_{n}\longrightarrow\epsilon+i0^{+})\Big)^{2},
 \label{e5.6}
\end{eqnarray}
where $n_{F}(x)=\frac{1}{e^{x/k_{B}T}+1}$ is the Fermi-Dirac
distribution function and $T$ denotes the equilibrium temperature.
Substituting electronic Green's function into Eq.(\ref{e5.6}) and
performing the numerical integration over wave vector through first
Brillouin zone, the results of optical absorption have been
obtained. Here, the contribution of both inter and intra band
transitions on the optical conductivity in Eq.(\ref{e5.6}) has been
considered.
\section{Results and conclusions}
We have obtained
the optical conductivity of the gapped biased
 undoped bilayer graphene for both stacking cases for polarization of electromagnetic wave
along $x$ direction as shown in Fig.(1).
We have implemented a tight binding model Hamiltonian including gap parameter
and bisa voltage. We have obtained the optical spectrum of
this structure by means of Green's function
approach so that the optical conductivity has been obtained by
calculating the electrical current
correlation function.
Using the band structures in Eqs.(\ref{e0.6},\ref{e0.6.1.1}), the electronic Green's function
is found based on Eq.(\ref{e0.61}) and afterwards the photon frequency dependence of
optical absorption is obtained by Eq.(\ref{e5.6}).
 In the obtaining the following numerical results,
the intralayer nearest neighbor hopping parameter ($t$) is set to 1.
Therefore the other parameters in the model Hamiltonian are expressed
as $\gamma/t$, $V/t$,.

The optical absorption of gapped simple stacked bilayer
graphene structure as a function of normalized frequency $\omega/t$ for different values of
gap parameters, namely $\Delta/t=0.3,0.6,0.9,1.2$ for half filling case
 $\mu/t=0$ are shown in Fig.(2).
For gap parameter cases $\Delta/t=0.3,0.6$,
optical absorption $\sigma(\omega)$ displays a Drude response or ballistic transport
\cite{grosso} at zero frequency limit
which arises from intraband transitions. Such infinite conductivity
originates from classical motion of electrons under long wave length of photons.
 This Drude response has been predicted under
Dirac approximation for gapless graphene\cite{stauber1}.
Upon more increasing gap parameter, i.e. $\Delta/t=0.9,1.2$, the band gap width
 in density of states increases so that Drude wight coefficient in optical conductivity
 goes to zero as shown in Fig.(2). Insulating behavior in bilayer graphene becomes more remarkable with increasing
gap parameter and consequently the Drude wight vanishes for normalized gap parameters 0.9,1.2.
Interband transition contributes to the optical conductivity at higher frequencies
 with increasing gap parameter.
 It can be understood from this fact
that the interband transition requires photons with more frequencies when the band gap in density of states increases.
In other words, the characteristic frequency position of peak in optical absorption tends to higher frequencies with gap parameter $\Delta$ according to
 Fig.(2).
 Upon increasing frequency above the peak position, the optical conductivity reduces
 so that it vanishes
at $\omega/t\approx7.0$ for all values of gap parameter.

In Fig.(3), we have plotted optical absorption of undoped AA stacked
bilayer graphene as a function of frequency for the various
normalized temperature, namely $k_{B}T/t=0.04,0.05,0.063,0.07,0.075$
 for bias voltage $V/t=0.0$ at $\Delta/t=0.0$.
Due to increase of transition rate of electrons with temperature, an increasing behavior for
 $\sigma(\omega)$ is clearly observed on the whole range of frequency. Both inter and intraband transition rates raises with temperature so that
the Drude weight and height of peak increase in optical spectrum of bilayer graphene in Fig.(3).
The frequency position of peaks in optical absorption is independent of
temperature and is around $\omega/t\approx2.5$

Also the frequency dependence of optical spectrum goes to zero in the frequency region
above normalized value 7.0 for all values of temperature.

 In Fig.(4), we present in-plane
optical conductivity of the undoped AA stacked bilayer graphene versus normalized
frequency ($\omega/t$),
 for different values of bias voltage,
namely $V/t=0.0,0.5,1.0,1.3$ for
 fixed temperature $k_{B}T/t=0.06$ and $\Delta/t=0.0$.
 For high values of bias voltages $V/t=1.0,1.3$,
the dynamical conductivity displays a Drude weight at low frequency due to
intraband transitions. However the optical conductivity has no
 Drude weight for lower bias voltages $V/t=0.0,0.5$.
 The emergence of Drude weight has been predicted by
using Dirac cone approximation\cite{tarbetArxiv}.
The optical conductivity curves in Fig.(4) have a broad peak at finite non zero frequency
for all values of bias voltage. Such peak arises from interband electronic transition
so that the frequency position of peaks is around $\omega/t\approx2.75$.
However the height of peaks reduces with bias voltage as shown in Fig.(4).
At fixed frequencies below normalized value 2.0, the optical conductivity decreases with
bias voltage. For bias voltage values $V/t=0.0,0.5$, the optical absorption gets the zero values
in frequency region $0<\omega/t<0.75$. In fact electromagnetic wave is not absorbed by bilayer structure
in this frequency
region for $V/t=0.0,0.5$.

 We have also studied the optical conductivity of AB stacked bilayer graphene.
In Fig.(5), we have plotted the frequency dependence of
 optical conductivity of gapped bernal stacked bilayer
graphene structure as a function of normalized frequency $\omega/t$ for different values of
gap parameters for half filling case in the absence of bias voltage.
For gap parameter cases $\Delta/t=0.3$,
optical absorption $\sigma(\omega)$ displays a Drude response or ballistic transport
\cite{grosso} at zero frequency limit
which arises from intraband transitions.
Upon more increasing gap parameter, i.e. $\Delta/t=0.6,0.9,1.2$, the band gap width
 in density of states increases so that Drude wight coefficient in optical conductivity
 goes to zero as shown in Fig.(5). Insulating behavior in bilayer graphene becomes more remarkable with increasing
gap parameter and consequently the Drude wight vanishes for normalized gap parameters 0.6,0.9 and
1.2.
Interband transition contributes to the optical conductivity at higher frequencies
 with increasing gap parameter.
 It can be understood from this fact
that the interband transition requires photons with more frequencies when the band gap in density of states increases.
In other words, the characteristic frequency position of optical absorption tends to higher frequencies with gap parameter $\Delta$ according to
 Fig.(5).
 Upon increasing frequency above the peak position, the optical conductivity reduces
 so that it vanishes
at $\omega/t\approx7.0$ for all values of gap parameter.
The comparison between Figs.(2,5) shows the Drude weight for optical conductivities of simple stacked bilayer graphene
is higher than that of bernal bilayer graphene at gap parameter $\Delta/t=0.3$

The optical absorption of undoped unbiased AB stacked bilayer grapehene as a function
of frequency for the various normalized temperature has been plotted in Fig.(6).
The gap parameter $\Delta$ has been assumed to be zero.
Due to increase of transition rate of electrons with temperature, an increasing behavior for
 $\sigma(\omega)$ is clearly observed on the whole range of frequency.
Both inter and intraband transition rates raises with temperature so that
the Drude weight and height of peak increase in optical spectrum of bilayer graphene in Fig.(6).
Also the frequency dependence of optical spectrum goes to zero in the frequency region
above normalized value 7.0 for all values of temperature.

 In Fig.(7), we present in-plane
optical conductivity of the undoped AB stacked bilayer graphene versus normalized
frequency ($\omega/t$),
 for different values of bias voltage for
 fixed temperature $k_{B}T/t=0.06$ and $\Delta/t=0.0$.
 For high values of bias voltages $V/t=1.0,1.3$,
the dynamical conductivity displays a Drude weight at low frequency due to
intraband transitions. However the optical conductivity has no
 Drude weight for lower bias voltages $V/t=0.0,0.5$.
 The optical conductivity curves in Fig.(7) have a broad peak at finite non zero frequency
for all values of bias voltage. Such peak arises from interband electronic transition
so that the frequency position of peaks moves to higher frequencies with bias voltage.
Moreover the height of peaks reduces with bias voltage as shown in Fig.(7).
At fixed frequencies below normalized value 1.5, the optical conductivity decreases with
bias voltage. For bias voltage values $V/t=0.0,0.5$, the optical absorption gets the zero values
in frequency region $0<\omega/t<0.5$. In fact electromagnetic wave is not
 absorbed by bilayer structure
in this frequency
region for $V/t=0.0,0.5$.

Finally we have compared the behaviors of optical conductivities of simple and bernal stacked
bilayer graphenes at fixed gap parameter $\Delta/t=0.9$ for $V/t=1.0$ at $k_{B}T/t=0.07$.
The frequency position of broad peak in optical conductivity of bernal stacking locates at
higher
frequency rather than that in optical spectrum of simple stacked bilayer graphene.
Moreover the height of peak in optical spectrum of bernal stacked bilayer graphene is more
than that of simple stacking case.
Also the Drude weight in optical spectrum of simple stacked case gets higher value in comparison
with bernal stacked bilayer graphene case.

The optical conductivity of bilayer graphene is in agreement with 
the theoretical study of optical conductivity of bilayer graphene under Dirac cone approximation
\cite{tarbetArxiv}. The Drude weight in optical conductivity of bilayer graphene
has been obtained in our results in agreement with theoretical work \cite{tarbetArxiv}.
The effects of inter and intrasite interations between electrons on the physical properties
of the system such as excitonic corrections of optical properties can be studied using 
GW method. However the study of electronic interations in bilayer graphene is out of scope 
of the present manuscript.

\section{Conclusions}
In summary, we have studied the optical properties of
 bilayer grapehene in the presence of 
bias voltage and gap parameter. The effects of stacking type on the frequency dependence of
dynamical conductivity have been studied. Also we have investigated the effects of 
temperature on the conductivity.
The Drude weight due to intraband transition vanishes upon increasing gap parameter.
However the increase of temperature increases the Drude weight for stacking types of bilayer
graphene. The comparison between simple and bernal bilayer graphene shows the Drude weight 
in AA stacked bilayer graphene is higher than the Drude weight in AB one.

\begin{figure}
\begin{center}
\epsfxsize=0.8\textwidth
\includegraphics[width=15.cm]{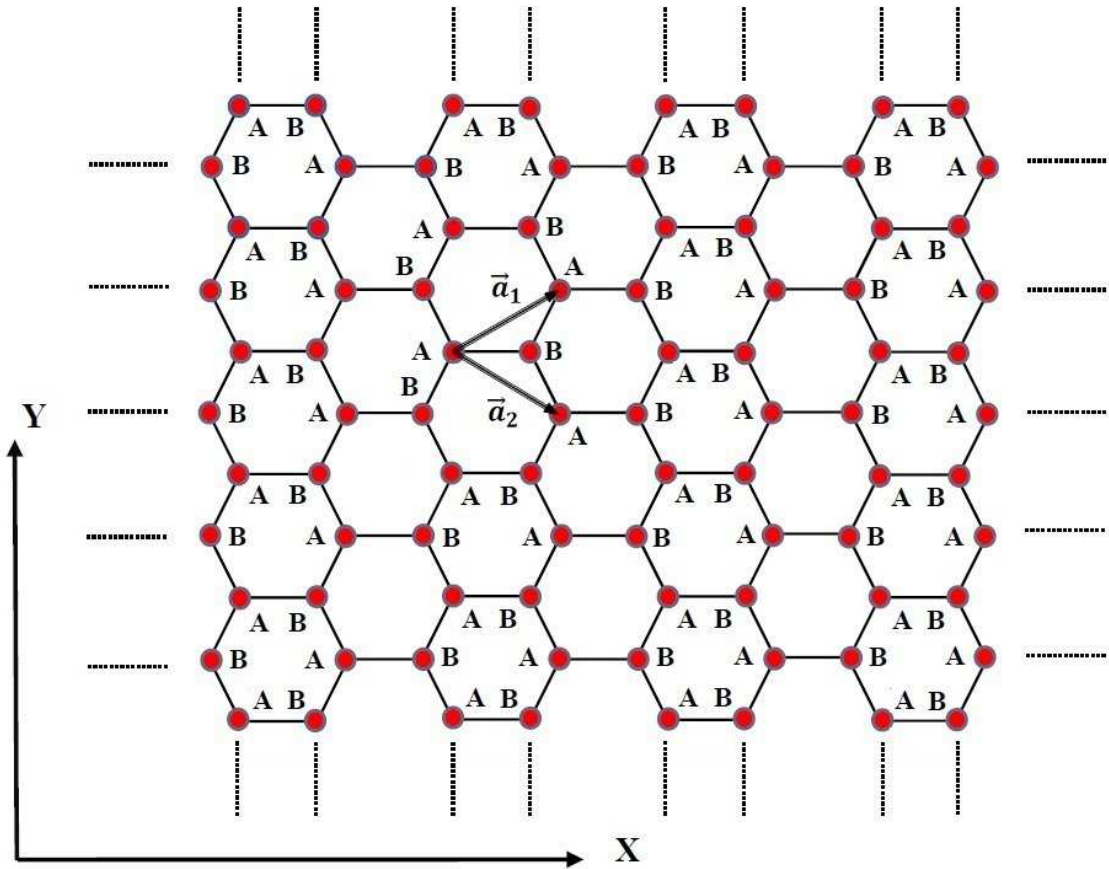}
\small
\begin{flushleft}
\caption{Crystal structure of honeycomb lattice with two different sublattices.
${\bf a}_{1}$ and ${\bf a}_{2}$ are the primitive unit cell vectors.}
\end{flushleft}
\end{center}
\end{figure}
\begin{figure}
\begin{center}
\epsfxsize=0.8\textwidth
\includegraphics[width=15.cm]{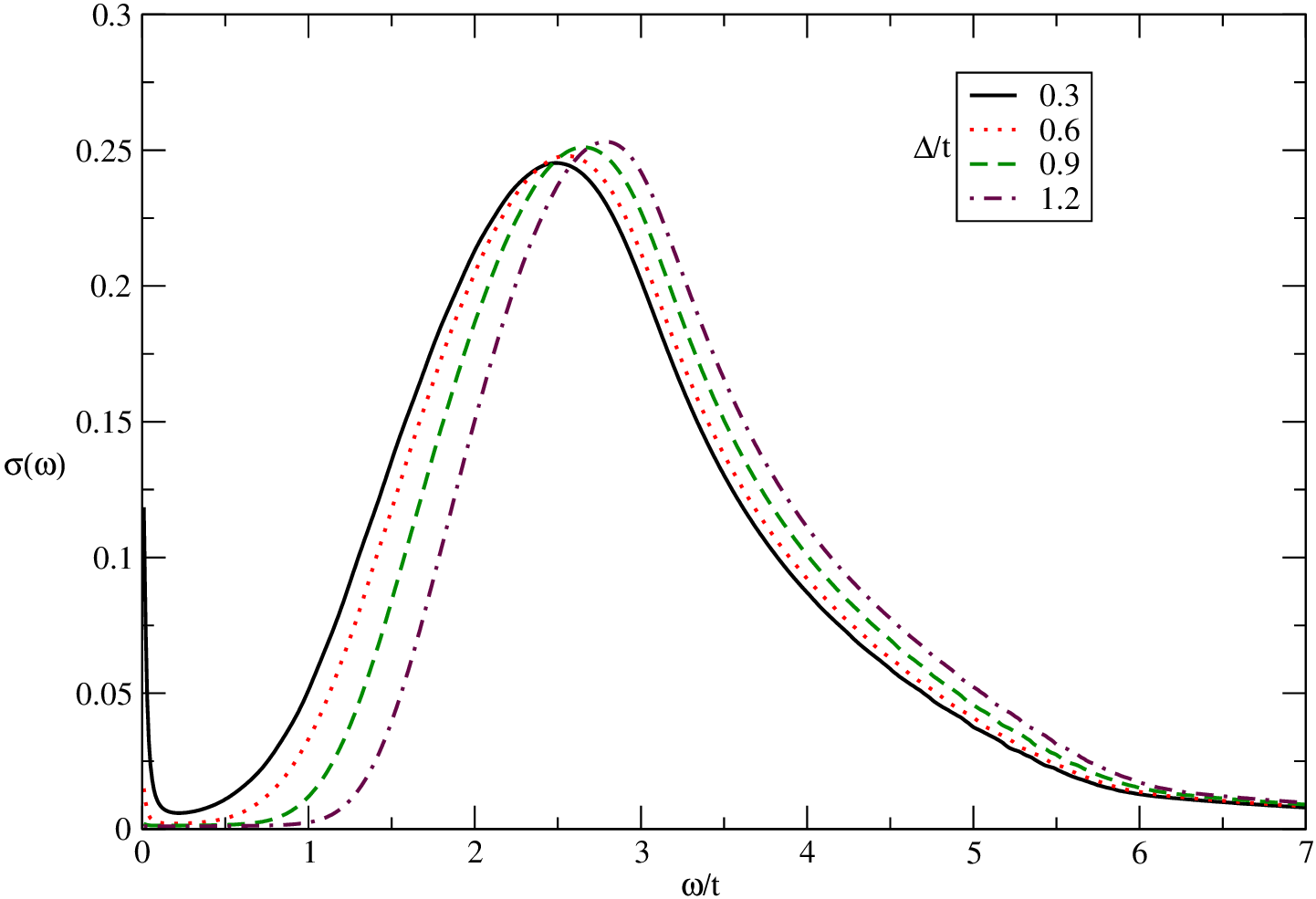}
\small
\begin{flushleft}
\caption{Optical conductivity ($\sigma(\omega)$) of unbiased AA
 stacked bilayer graphene as a function of normalized frequency $\omega/t$
for various amounts of gap parameter $\Delta/t_{\Arrowvert}$ for fixed temperature
$k_{B}T/t_{\Arrowvert}=0.06$ at zero value for chemical potential.}
\end{flushleft}
\end{center}
\end{figure}
\begin{figure}
\begin{center}
\epsfxsize=0.8\textwidth
\includegraphics[width=15.cm]{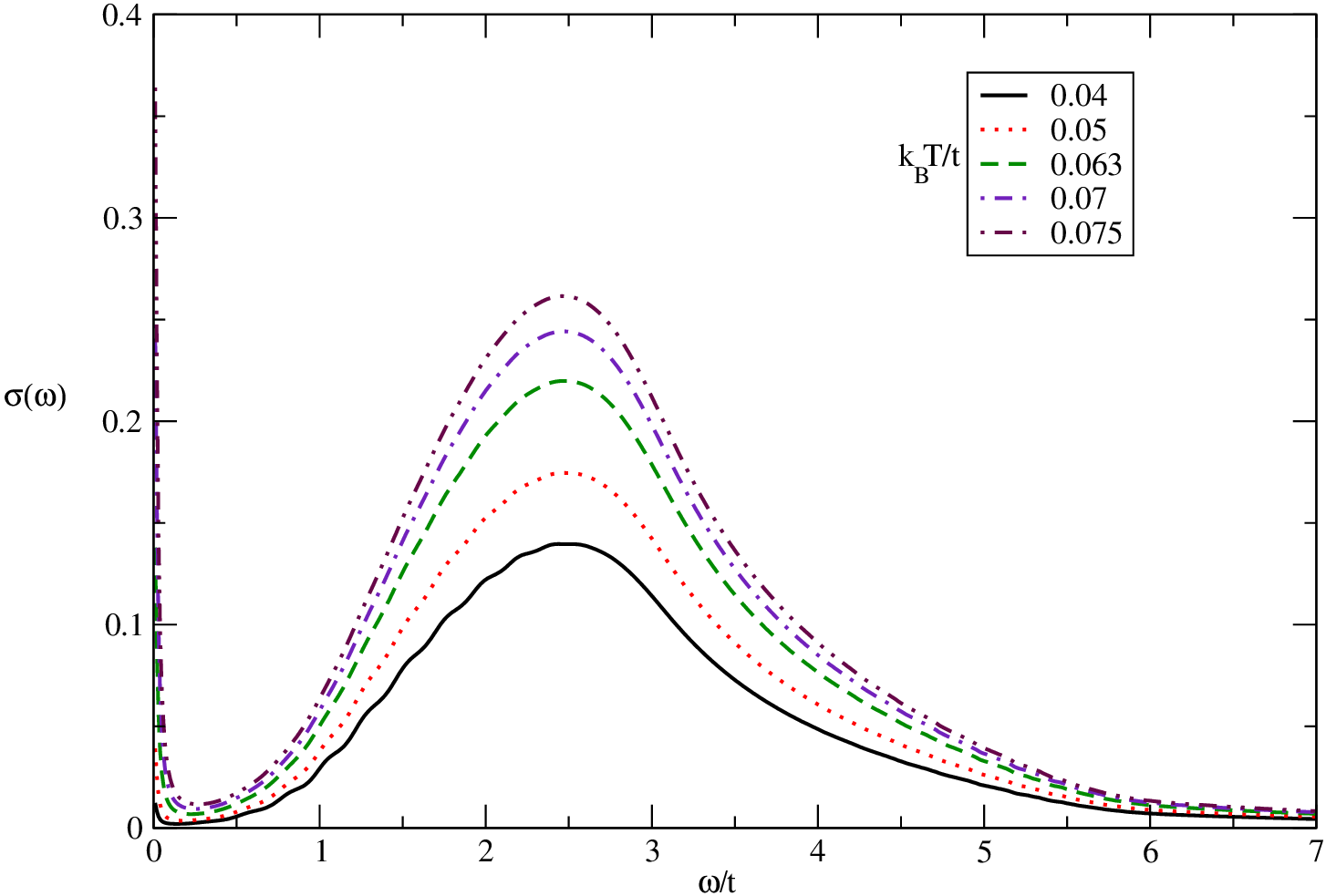}
\small
\begin{flushleft}
\caption{Optical conductivity ($\sigma(\omega)$) of unbiased AA
 stacked bilayer graphene as a function of normalized frequency $\omega/t$
for various amounts of normalized temperature $k_{B}T/t_{\Arrowvert}$ at fixed zero
gap parameter. The chemical potential has been assumed to be zero.}
\end{flushleft}
\end{center}

\end{figure}
\begin{figure}
\begin{center}
\epsfxsize=0.8\textwidth
\includegraphics[width=15.cm]{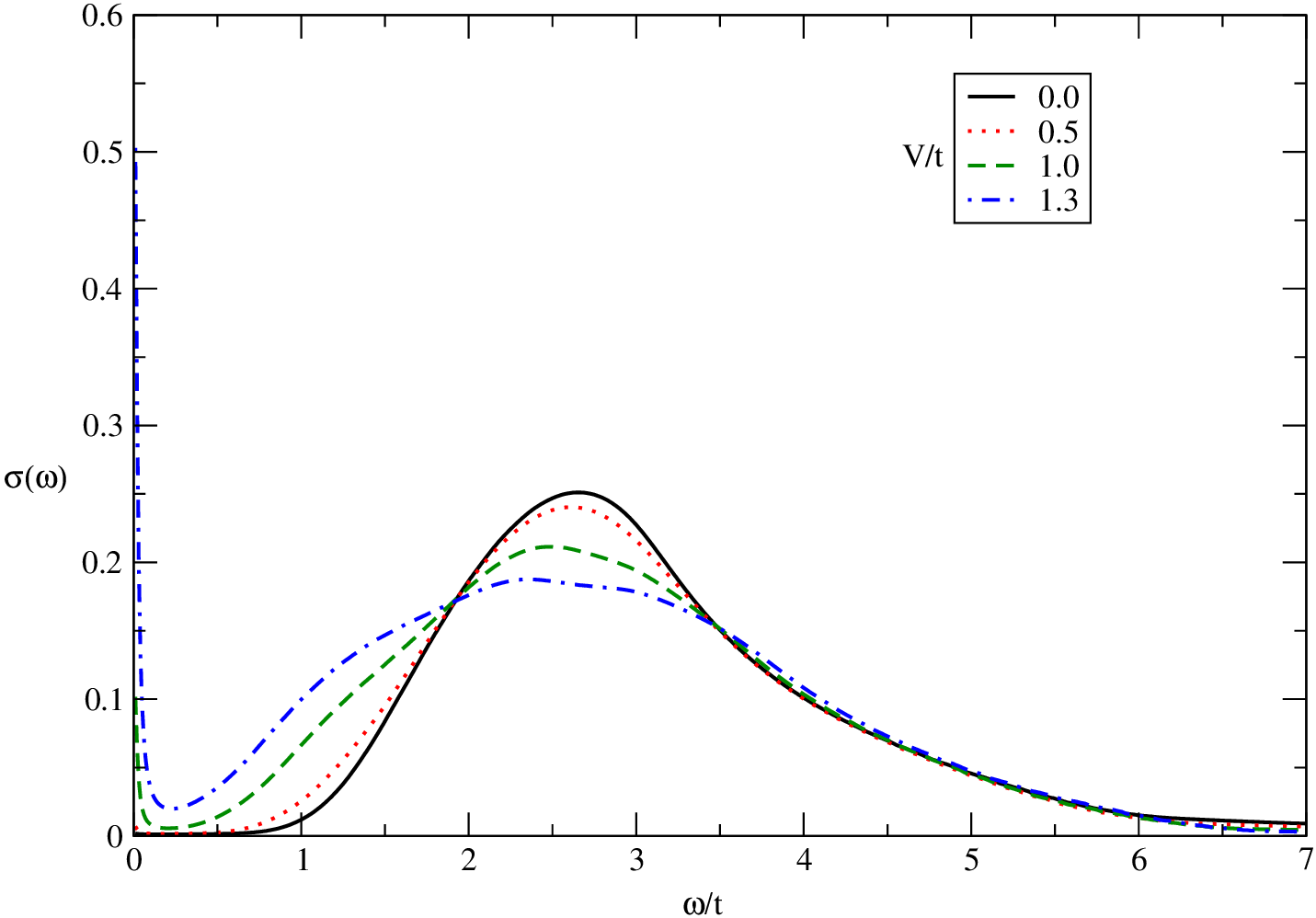}
\small
\begin{flushleft}
\caption{Optical conductivity ($\sigma(\omega)$) of biased undoped AA
 stacked bilayer graphene as a function of normalized frequency $\omega/t$
for various amounts of normalized bias voltage $V/t_{\Arrowvert}$ at fixed zero
gap parameter. The normalized temperature has been assumed to be $k_{B}T/t=0.06$.}
\end{flushleft}
\end{center}
\end{figure}
\begin{figure}
\begin{center}
\epsfxsize=0.8\textwidth
\includegraphics[width=15.cm]{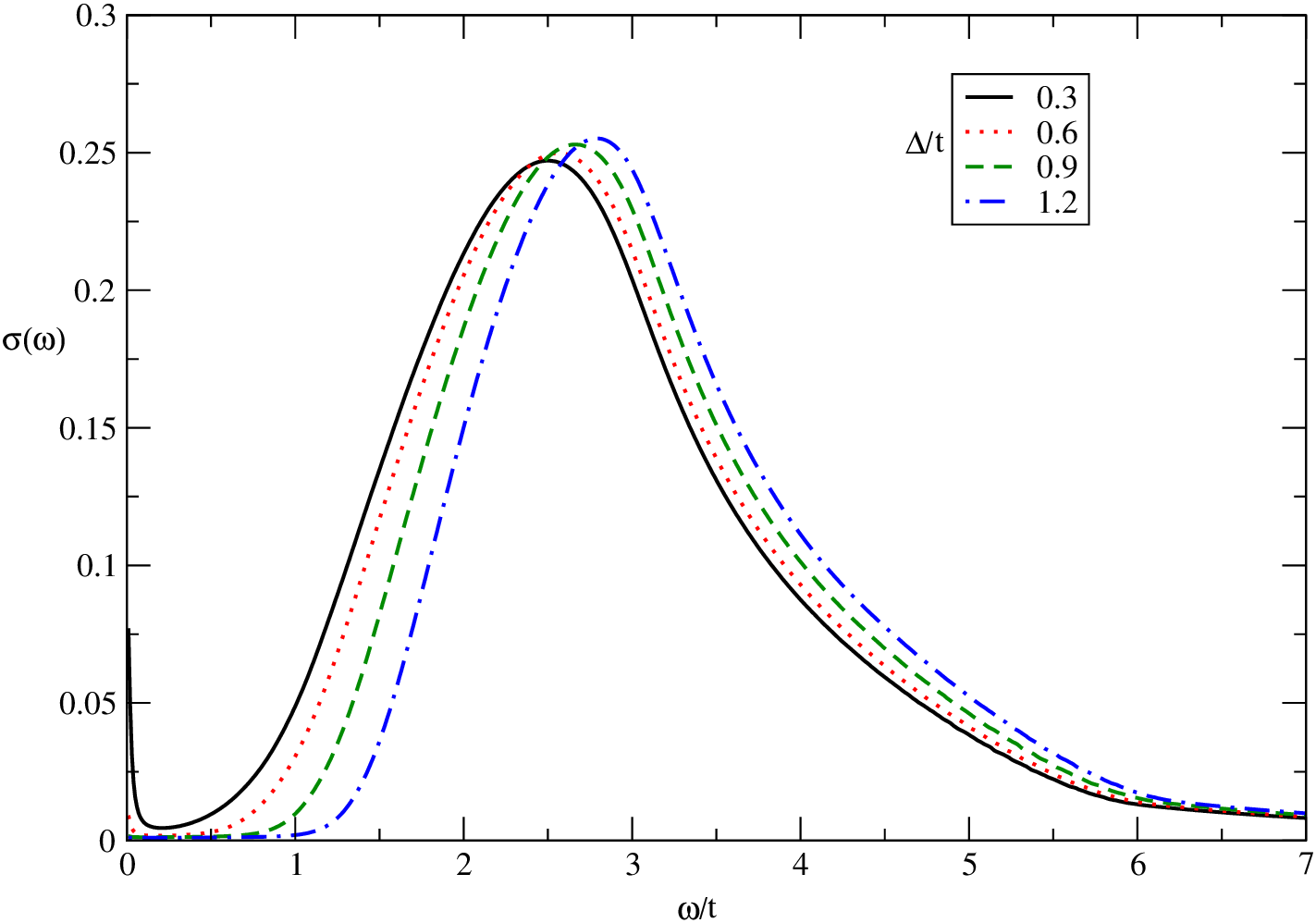}
\small
\begin{flushleft}
\caption{Optical conductivity ($\sigma(\omega)$) of unbiased AB
 stacked bilayer graphene as a function of normalized frequency $\omega/t$
for various amounts of gap parameter $\Delta/t_{\Arrowvert}$ for fixed temperature
$k_{B}T/t_{\Arrowvert}=0.06$ at zero value for chemical potential.}
\end{flushleft}
\end{center}
\end{figure}
\begin{figure}
\begin{center}
\epsfxsize=0.8\textwidth
\includegraphics[width=15.cm]{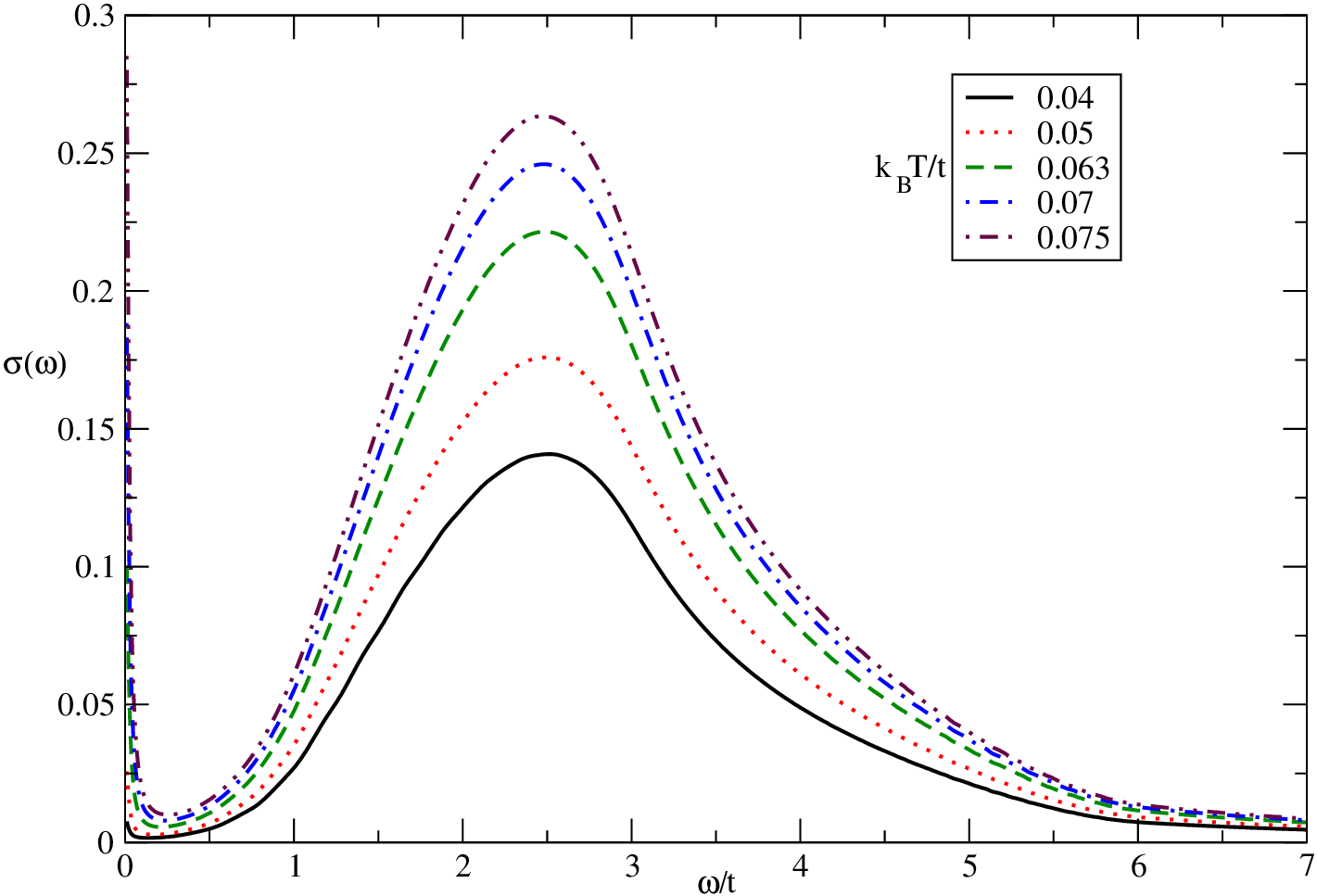}
\small
\begin{flushleft}
\caption{Optical conductivity ($\sigma(\omega)$) of unbiased AB
 stacked bilayer graphene as a function of normalized frequency $\omega/t$
for various amounts of normalized temperature $k_{B}T/t_{\Arrowvert}$ at fixed zero
gap parameter. The chemical potential has been assumed to be zero.}
\end{flushleft}
\end{center}
\end{figure}

\begin{figure}
\begin{center}
\epsfxsize=0.8\textwidth
\includegraphics[width=15.cm]{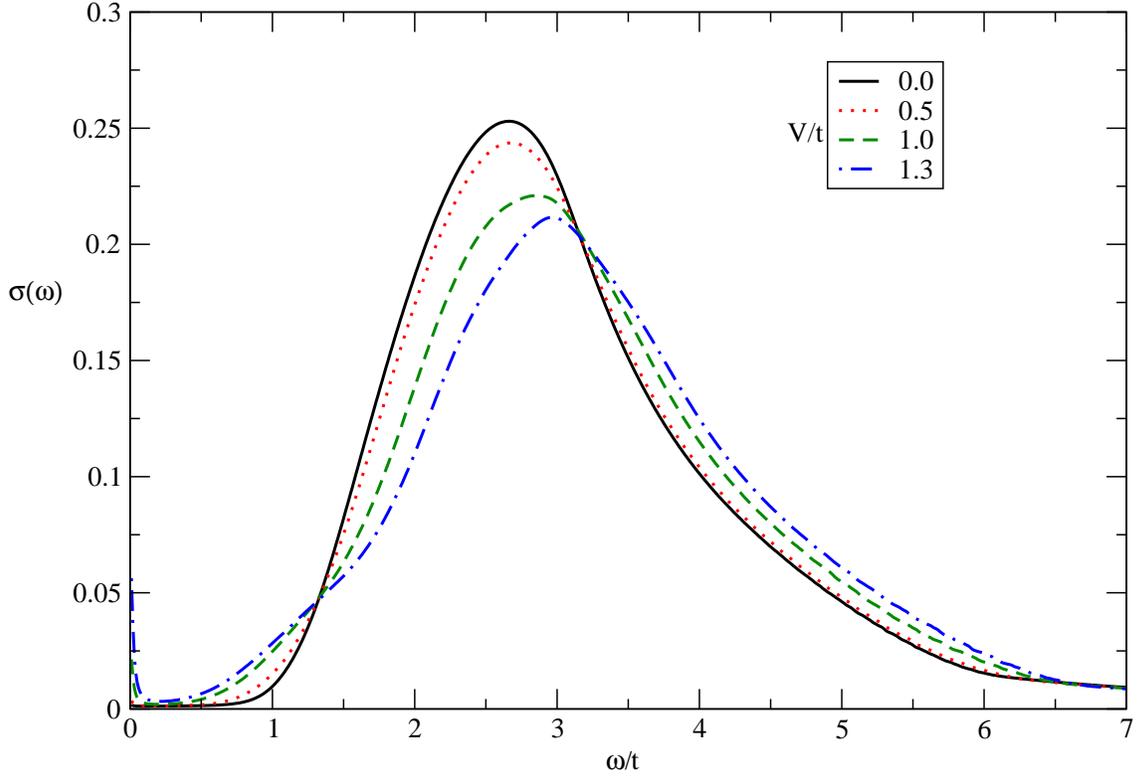}
\small
\begin{flushleft}
\caption{Optical conductivity ($\sigma(\omega)$) of biased undoped AB
 stacked bilayer graphene as a function of normalized frequency $\omega/t$
for various amounts of normalized bias voltage $V/t_{\Arrowvert}$ at fixed zero
gap parameter. The normalized temperature has been assumed to be $k_{B}T/t=0.06$.}
\end{flushleft}
\end{center}
\end{figure}

\begin{figure}
\begin{center}
\epsfxsize=0.8\textwidth
\includegraphics[width=15.cm]{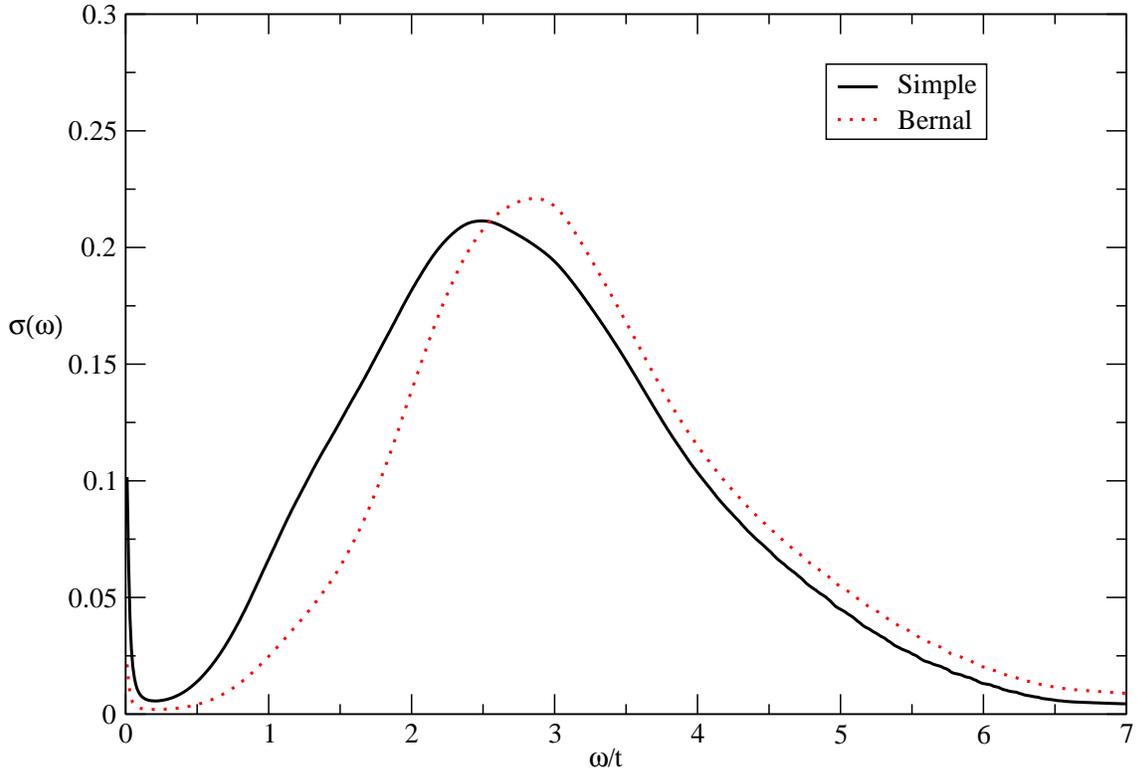}
\small
\begin{flushleft}
\caption{The comparison between optical conductivities ($\sigma(\omega)$) of simple (AA)
 and bernal (AB)
 stacked bilayer graphene as a function of normalized frequency $\omega/t$ for $k_{B}T/t=0.07$.
The normalized bias voltage has been fixed at $V/t_{\Arrowvert}=1$.
The normalized gap parameter has been assumed to be $\Delta/t=0.9$.}
\end{flushleft}
\end{center}
\end{figure}

\newpage

\end{document}